\documentclass[10pt]{iopart}
\usepackage{iopams,graphicx,color}  

\begin{document}

\title{Spin current compensation from competing magnon modes in ferrimagnets}

\author{Adam B. Cahaya}

\address{Department of Physics, Faculty of Mathematics and Natural Sciences, Universitas Indonesia, Depok 16424, Indonesia}
\ead{adam@sci.ui.ac.id}
\vspace{10pt}

\begin{abstract}
We investigate thermal spin pumping in gadolinium iron garnet (GdIG), focusing on the mode-resolved dynamics of antiferromagnetic magnons and their impact on spin and heat transport. Antiferromagnets support both right-handed and left-handed magnon modes, which we treat as positive and negative frequency branches, analogous to electrons and holes in semiconductors. Using a two-sublattice model with a minimal exchange interaction scheme, we derive the temperature-dependent spin-wave spectrum and evaluate the associated thermal spin pumping coefficients. Our analysis reveals that the competition between left- and right-handed modes gives rise to a compensation temperature, where the net thermally generated spin current vanishes.
Importantly, we show that this compensation point does not necessarily coincide with the crossing of magnon dispersion branches. While previous research considers a detailed microscopic model including all magnetic sublattices and exchange couplings, our approach demonstrates that key features of mode-resolved spin transport can be captured by a simplified and analytically transparent model. These findings advance the understanding of spin-caloritronic phenomena in ferrimagnets and offer new perspectives for the design of chiral magnon-based spintronic devices.
\end{abstract}

%
%
%
%
\ioptwocol

\section{Introduction}

The exploration of fundamental quasiparticles has driven significant advancements in our understanding of quantum mechanics and condensed matter physics. Among these quasiparticles, magnons, the quanta of spin waves, play a pivotal role in both theoretical frameworks and practical applications in spintronics and quantum information science \cite{Lebrun2020,Cornelissen2015}. In particular, antiferromagnetic magnons, which arise in systems with alternating spin orientations, exhibit rich physical properties due to their multiple propagation modes and interactions.

Building on this concept, researchers have introduced magnons with distinct handedness, referring to the direction of spin precession: left-handed and right-handed. This property emerges due to the underlying magnetic symmetry and adds a new degree of freedom in magnonic systems, enabling chirality-based manipulation of spin waves \cite{Kanj2023}. Thermal spin pumping has become a promising technique for detecting and utilizing these chiral magnon modes \cite{Siddiqui2021}. By converting thermally excited spin waves into measurable electrical signals, this method provides an indirect but powerful probe of the magnonic spectrum.

Rare-earth iron garnets (RIGs) are particularly well-suited for studying such effects due to their complex magnetic structures and ultra-low magnetic damping \cite{Seo2024}. The three magnetic sublattices — tetrahedral Fe$^{3+}$, octahedral Fe$^{3+}$, and dodecahedral R$^{3+}$ sites — give rise to multiple magnonic branches \cite{BarkerBauer2016,Huang2024}. Tuning the rare-earth ion content enables precise control of the spin-wave modes \cite{Zhang2023}. Moreover, strong spin-orbit coupling effects, such as those from the Dzyaloshinskii-Moriya interaction and the inverse Faraday effect, further enhance their potential for spintronic applications \cite{Cheng2016}.
{
A particularly intriguing feature in thermal spin pumping in ferrimagnets is the presence of two compensation temperatures. The first corresponds to the compensation of net magnetic moment, which is well-established. The second, at a lower temperature, has been proposed to result from the competition between left-handed and right-handed magnons. In Ref.\cite{Li2022}, this lower compensation point was observed to coincide with the crossing of the $|\boldsymbol\omega_R|$ and $|\boldsymbol\omega_L|$ spectra, suggesting a connection between spectral symmetry and thermal spin current reversal. However, the separate contributions of left- and right-handed modes to thermal spin pumping have not yet been quantitatively evaluated.

In this manuscript, we address this gap by explicitly calculating the contributions of each mode using a two-magnetic-lattice model, allowing mode-resolved analysis of thermal spin currents. Our model reveals that the lower compensation temperature does not align with the spectral crossing of $|\boldsymbol\omega|$, and conversely, that a spectral crossing can occur even when thermal spin pumping remains finite. These findings question the assumption that compensation behavior is directly linked to spectral symmetry and highlight the importance of a detailed, mode-specific approach.

\begin{figure}[t]
\includegraphics[width=\columnwidth]{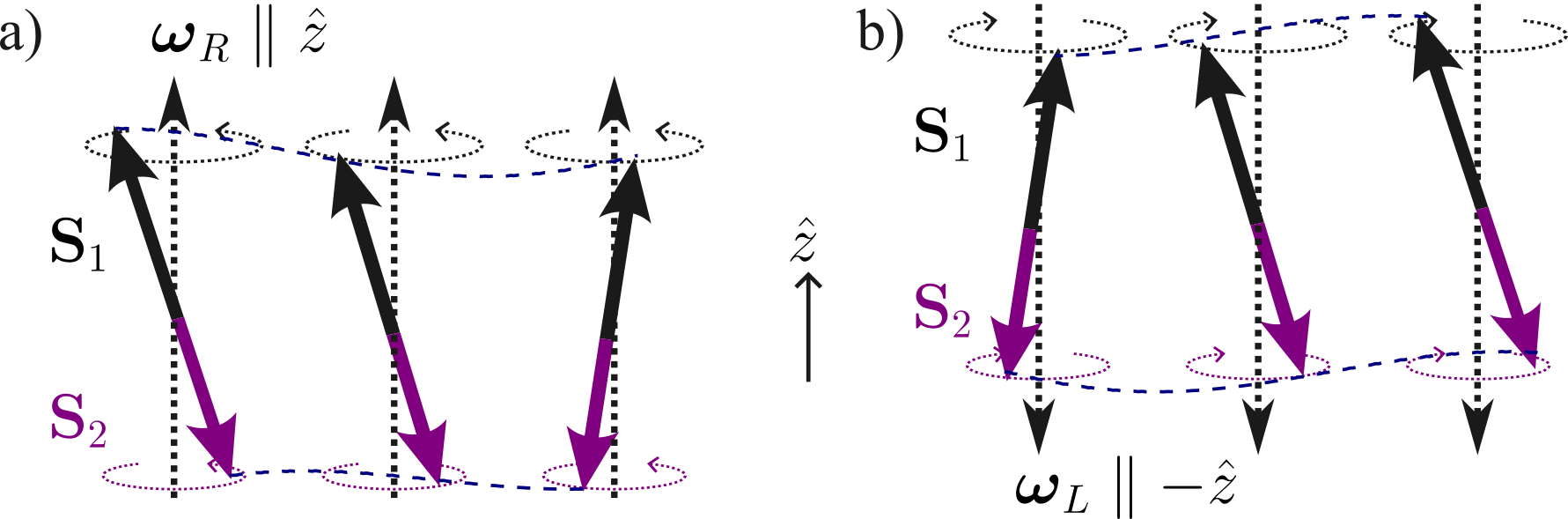}
\caption{(a) Right-handed magnon precessing with radial frequency $\omega_R$ in counter clockwise direction around $+z$ axis and left-handed magnon precessing with radial frequency $\omega_L$ in clockwise direction.}
\label{Figantiferromagnon}
\end{figure}

To interpret these results, we treat right-handed and left-handed magnon modes as positive and negative frequency branches, respectively. This treatment draws an analogy to electronic transport in semiconductors, where electrons and holes contribute oppositely to current due to their energy band positions. By applying this analogy to spin transport, we develop a formalism in which the net thermal spin current arises from the imbalance between left- and right-handed magnon populations, governed by their respective dispersion relations and thermal distributions.

While Ref.~\cite{Li2022} employs a comprehensive microscopic model of GdIG that includes detailed exchange interactions among all magnetic sublattices, our approach is deliberately simplified. We adopt a two-magnetic-lattice model incorporating only the essential Fe–Fe and Fe–Gd exchange interactions. This reduced model is analytically tractable and transparent, yet sufficiently rich to capture key features of the mode-resolved magnon dynamics, including the emergence of compensation temperatures and the role of spectral symmetry in spin transport.

The remainder of this paper is organized as follows: In Section~\ref{sec:methods}, we outline the formalism for calculating spin and heat currents based on the spin-mixing conductance and magnon dynamics, and derive analytic expressions for the spin transport coefficients $\mathcal{L}_0$ and $\mathcal{L}_1$. In Section~\ref{sec:results}, we apply a two-sublattice model of GdIG to compute the temperature-dependent magnon spectrum and investigate the contributions of right- and left-handed magnons to thermal spin pumping. We identify the compensation conditions and provide a quantitative analysis of mode-resolved spin transport. Finally, we summarize our conclusions and discuss future perspectives in Section~\ref{sec:conclusion}.

\section{Methods}\label{sec:methods}}
In thermal spin pumping, spin $J_s$ and heat $J_Q$ currents are transferred from a magnetic material, such as rare-earth iron garnet, to a non-magnetic metal, such as Pt. This spin current is then converted into an electrical signal via the inverse spin Hall effect. At the interface, $J_s$ and $J_Q$  coupled and can be expressed through a linear matrix relation involving the driving forces: spin accumulation $\boldsymbol{\mu}_s$ and temperature difference $\Delta T$, analogous to the thermoelectric effect:
\begin{equation}
\left[\begin{array}{c}
J_s\\
J_Q
\end{array} \right]
=\left[\begin{array}{cc}
\mathcal{L}_0 & \mathcal{L}_1\\
\mathcal{L}_1 & \mathcal{L}_2
\end{array} \right]
\left[\begin{array}{c}
-\hat{\textbf{m}}\cdot\boldsymbol{\mu}_s\\
-\Delta T/T
\end{array} \right].
\label{EqMatriksSpinHeat}
\end{equation}
Here, $\hat{\textbf{m}}=\hat{z}S_z/|S_z|$ is the average direction of the magnetization. $\mathcal{L}_i$ is analogous to thermoelectric transport coefficients. $\mathcal{L}_0$ represents the interface spin injection conductance, while $\mathcal{L}_2$ denotes the heat conductance. Since the heat conductance at the interface is generally dominated by Kapitza resistance \cite{Cahaya2014}, our discussion is focused on $\mathcal{L}_0$ and $\mathcal{L}_1$. The ratios $\mathcal{L}_1/\mathcal{L}_0$ and $\mathcal{L}_1/(\mathcal{L}_0T)$ correspond to the spin Peltier and spin Seebeck coefficients, respectively \cite{Cahaya2014}. 

The expression of $\mathcal{L}_0$ and $\mathcal{L}_1$ can be determined from total spin current transferred at the interface.
\begin{eqnarray}
J_s=& J_\mathrm{bf}+J_\mathrm{sp}+J_\mathrm{fl} , \label{Eq.Spincurrents}
\end{eqnarray}
where $J_\mathrm{sp}$ arises from the spin pumping
\begin{eqnarray}
J_\mathrm{sp}=G\left<\textbf{S}\times\dot{\textbf{S}}\right>\cdot\hat{z}.
\end{eqnarray}
$J_\mathrm{fl}=-\gamma\left<\textbf{S}\times\textbf{h}'\right>\cdot\hat{z}$ is spin backflow due to thermal noise. $J_\mathrm{fl}=-J_\mathrm{sp}(T-\Delta T)/T$ has been shown to balances $J_\mathrm{sp}$ when there is no temperature difference across interface \cite{PhysRevB.81.214418}.
\begin{eqnarray}
J_\mathrm{sp}+J_\mathrm{fl}=&\frac{\Delta T}{T}G\left<\textbf{S}\times\dot{\textbf{S}}\right>\cdot\hat{z}.
\end{eqnarray}
Lastly, $J_\mathrm{bf}$ arises from spin backflow due to spin accumulation polarized along the $z$-direction:
\begin{eqnarray}
J_{\mathrm{bf}}=&G\left<\textbf{S}\times\left(\textbf{S}\times\hat{z}\mu_s\right)\right>\cdot\hat{z}
=-G\left<S_x^2+S_y^2\right>\mu_s
\end{eqnarray} 
where $G$ is the spin-mixing conductance\cite{Jiao2013}. 

Using the Holstein-Primakoff transformation $S_+=S_x+ iS_y\simeq\sqrt{2S}a$, $S_-=S_x-iS_y\simeq\sqrt{2S}a^\dagger$, 
$S_z=S-a^\dagger a$, we obtain
\begin{eqnarray}
\mathcal{L}_0&=&-\frac{\partial J_s}{\partial\hat{\textbf{m}}\cdot\boldsymbol{\mu}_s}
=\frac{G|S_z|}{2S_z}\left<S_-S_++S_+S_-\right>\nonumber\\
&\simeq& 2G|S_z| \left<a^\dagger a\right>.
\end{eqnarray}
The expectation value of the bosonic field of the ferromagnetic spin wave $\omega=\omega_0+Dk^2$ gives
\begin{eqnarray}
\mathcal{L}_0
&=&2GS_z \int_{\omega_0}^\infty \frac{\mathrm{DOS}(\omega)d\omega}{e^{\beta\omega}-1}\nonumber\\
&=&\frac{GS_z}{2}\left(\frac{k_BT}{\pi D}\right)^{\frac{3}{2}}\mathrm{Li}_{\frac{3}{2}}\left(e^{-\beta\omega_0}\right), \label{Eq.L0}
\end{eqnarray}
where $\beta=(k_BT)^{-1}$, $\mathrm{Li}_n(x)$ is the polylogarithm function, with $\mathrm{Li}_n(1)=\zeta(n)$ reducing to the Riemann zeta function.
{The rest of the spin current in Eq.~\ref{Eq.Spincurrents} is contributes to} $\mathcal{L}_1$ 
\begin{eqnarray}
\mathcal{L}_1&=&\frac{\partial J_s}{\partial (\Delta T/T)}= G\left<\textbf{S}\times\dot{\textbf{S}}\right>\cdot\hat{z}
= 2GS_z \left<a^\dagger \omega a\right> \nonumber\\
&=& 2GS_z \int_0^\infty \frac{\omega\mathrm{DOS}(\omega)d\omega}{e^{\beta\omega}-1}\nonumber\\
&=&\frac{3GS_z\left(k_BT\right)^{\frac{5}{2}}}{4\left(\pi D\right)^{\frac{3}{2}}}\mathrm{Li}_{\frac{5}{2}}\left(e^{-\beta\omega_0}\right).
\end{eqnarray}

In YIG, the magnon gap $\omega_0=\gamma H$ can be non zero due to Zeeman effect \cite{Rodrigue1960}. Since $\lim_{x\to 0}\mathrm{Li}_{n}(x)\simeq x$, the magnetic field $H$ exponentially suppress thermal spin pumping \cite{Kikkawa2015}
$\mathcal{L}_1\propto e^{-\gamma H/k_BT}.$ 
The magnetization of YIG originates from five Fe$^{3+}$ ions distributed across two magnetic lattices. Two of these ions occupy tetrahedral Fe$^{3+}$ sites, while three are in octahedral sites. Due to their antiferromagnetic coupling, only one Fe$^{3+}$ spin contributes \cite{NEEL1964344}, as illustrated in Fig.~\ref{FigCoupling}a. When yttrium in YIG is substituted by lanthanides, R$^{3+}$ occupies the dodecahedral site, forming an additional magnetic lattice. This substitution leads to a more complex spin-wave spectrum compared to that of YIG. While significant computational efforts are underway to unravel these spectra \cite{Wang2020,Li2022}, determining thermal spin pumping from detailed spectra remains a formidable challenge \cite{Shen2019}. Additionally, conduction electrons interact differently with localized spin and orbital angular momentum \cite{Kondo1962}. To focus on competition of antiferromagnetic magnon due to ferrimagnetically coupled spins, we restrict our discussion to gadolinium iron garnet (GdIG), which exhibits zero orbital angular momentum.

\subsection{Statistical mechanics of GdIG.}

A model that provides considerable insight into thermal spin pumping in GdIG is the two-magnetic-lattice model \cite{Cahaya2022,Cahaya2022PLA}. In this model, the magnon spectrum of GdIG is determined by two magnetic lattices: $S^\mathrm{Gd}$, representing the total angular momentum of three Gd$^{3+}$ ions, and $S^\mathrm{Fe}$, representing the spin angular momentum of Fe$^{3+}$ ions. This model has been shown to describe both its low-lying spin wave spectrum \cite{Tinkham1961} and room-temperature thermal spin pumping \cite{Cahaya2022}.

\begin{figure}[t]
\includegraphics[width=\columnwidth]{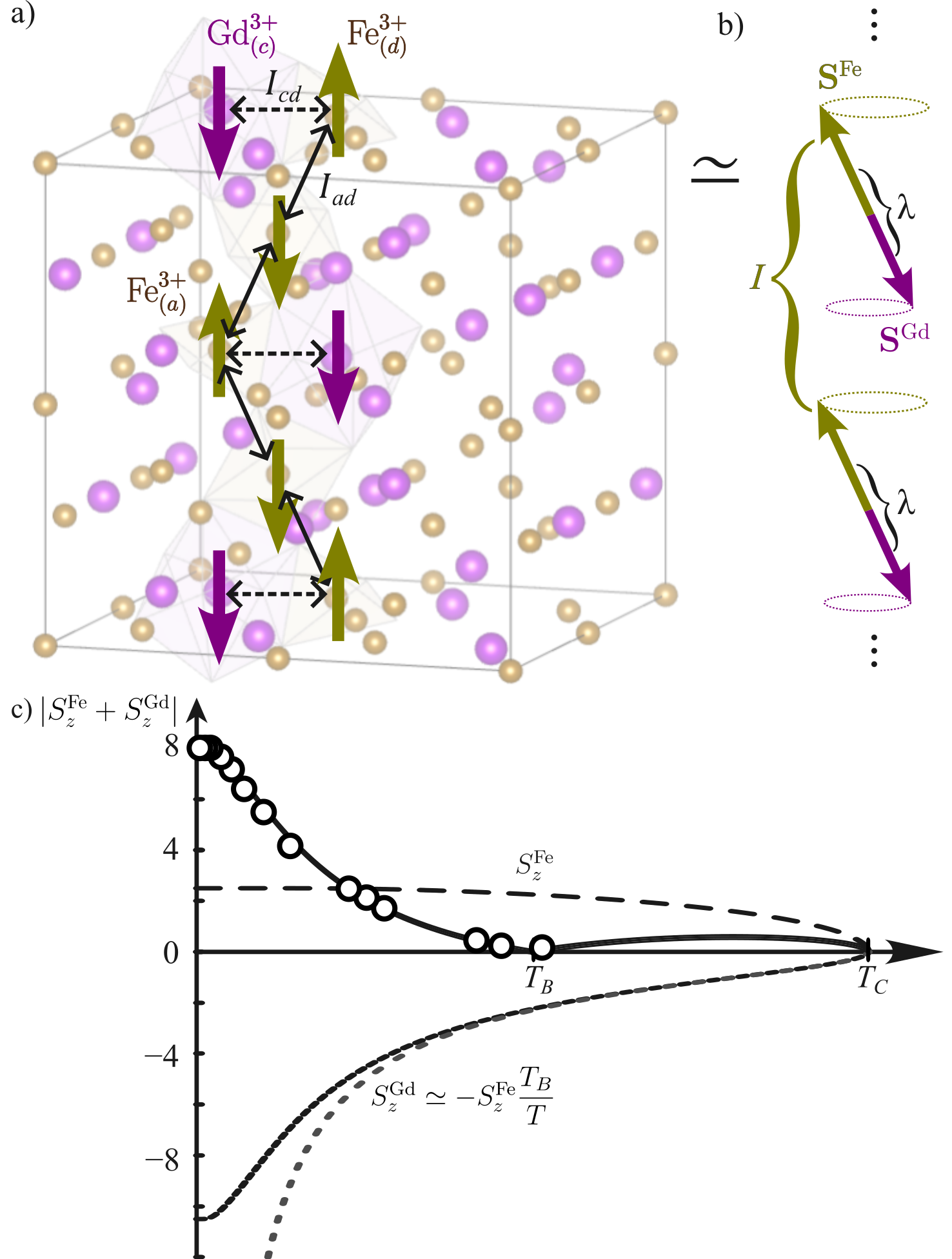}
\caption{(a) Exchange coupling between spins of Fe$^{3+}$, occupying tetrahedral and octahedral sites, and Gd$^{3+}$, occupying dodecahedral sites, in Gd$_3$Fe$_5$O${12}$ (GdIG) can be approximated by (b) the two-spin model with $\mathrm{S}^\mathrm{Fe}$ and $\mathrm{S}^\mathrm{Gd}$. (c) Temperature dependence of the $z$-component of spins in GdIG is in agreement to Ref.~\cite{Geller1961}. The total spin is zero at the compensation temperature $T_B$.}
\label{FigCoupling}
\end{figure}

We consider two isotropic exchange couplings: Fe$^{3+}$-Fe$^{3+}$ coupling, characterized by the exchange constant $I$, and Fe$^{3+}$-Gd$^{3+}$ coupling, characterized by the exchange constant $\lambda$, as illustrated in Fig.~\ref{FigCoupling}b. 
\begin{eqnarray}
\mathcal{H}=&-\frac{I}{2}\sum_{j} \textbf{S}^\mathrm{Fe}_{j}\cdot\textbf{S}^\mathrm{Fe}_{j+1}+\lambda\sum_{j} \textbf{S}^\mathrm{Gd}_{j}\cdot\textbf{S}^\mathrm{Fe}_{j} , \label{EqHint}
\end{eqnarray}
where the subscript $n$ indexes the Fe$^{3+}$ chain. The Fe-Gd coupling induces a paramagnetic response of $S^\mathrm{Gd}$ to the molecular field generated by $S^\mathrm{Fe}$. Gd-Gd coupling is neglected, as validated by the dominance of Fe-Fe coupling in the Curie temperature $T_C= IS^\mathrm{Fe}(S^\mathrm{Fe}+1)/(3k_B)+\mathcal{O}\left(\lambda^2/I\right)$, as derived in the \ref{SecAppendix}. The effect of Gd-Gd exchange, which is order of magnitudes smaller than $\lambda$ \cite{Gilleo1980}, is also estimated in the Appendix.
Fig.~\ref{FigCoupling}c illustrates the temperature dependence of the spins, obtained from the statistical average:
\begin{eqnarray}
S^\mathrm{Gd}_z&=&3S^\mathrm{Gd}B_{S^\mathrm{Gd}}\left(S^\mathrm{Gd}\frac{-\lambda S^\mathrm{Fe}_z}{k_BT}\right)
\simeq
-\frac{S^\mathrm{Gd}(S^\mathrm{Gd}+1) \lambda S^\mathrm{Fe}_z}{k_BT},\nonumber\\
S^\mathrm{Fe}_z&=&S^\mathrm{Fe}B_{S^\mathrm{Fe}}\left(S^\mathrm{Fe}\frac{I S^\mathrm{Fe}_z-\lambda S^\mathrm{Gd}_z}{k_BT}\right),
\label{Eq_Sgd}
\end{eqnarray} 
where $B_J(x)$ is the Brillouin function. A compensation temperature occurs at \cite{Cahaya2022}
\begin{equation}
T_B=k_B^{-1}\lambda S^\mathrm{Gd}(S^\mathrm{Gd}+1).
\end{equation}
For GdIG, $T_B\simeq 0.5 T_C$, $\lambda\simeq 0.1 I$.

Compared to the detailed microscopic approach taken in Ref.~\cite{Li2022}, which models all magnetic sublattices and their exchange interactions explicitly, our two-magnetic-lattice model is highly simplified. It includes only the dominant Fe–Fe and Fe–Gd isotropic exchange couplings and neglects orbital contributions, dipolar effects, and anisotropies. This simplification allows for an analytically tractable framework that nonetheless reproduces key features of the magnonic structure and spin transport in GdIG, including the temperature dependence of compensation behavior and mode competition. 

We emphasize that the omission of magnetic anisotropy and dipolar interactions is a deliberate choice to isolate the role of exchange-driven mode competition. Fig.~\ref{FigCoupling}c illustrates the agreement of our minimal model with experiment in Ref.~\cite{Geller1961}, this show that isotropic exchange interactions alone are sufficient to reproduce the temperature dependence of magnetization in GdIG. Furthermore, while spin–orbit coupling is typically significant in rare-earth systems, gadolinium ions (Gd$^{3+}$) possess a half-filled 4$f^7$ shell with zero orbital angular momentum, effectively suppressing spin–orbit effects. This justifies modeling GdIG as a spin-only system for the purposes of studying thermally induced magnon transport. Although the model does not aim to replicate the full magnon spectrum complexity of GdIG, it captures the essential physics relevant to thermal spin pumping and mode-resolved contributions with minimal computational effort.

\begin{figure}[t]
\includegraphics[width=\columnwidth]{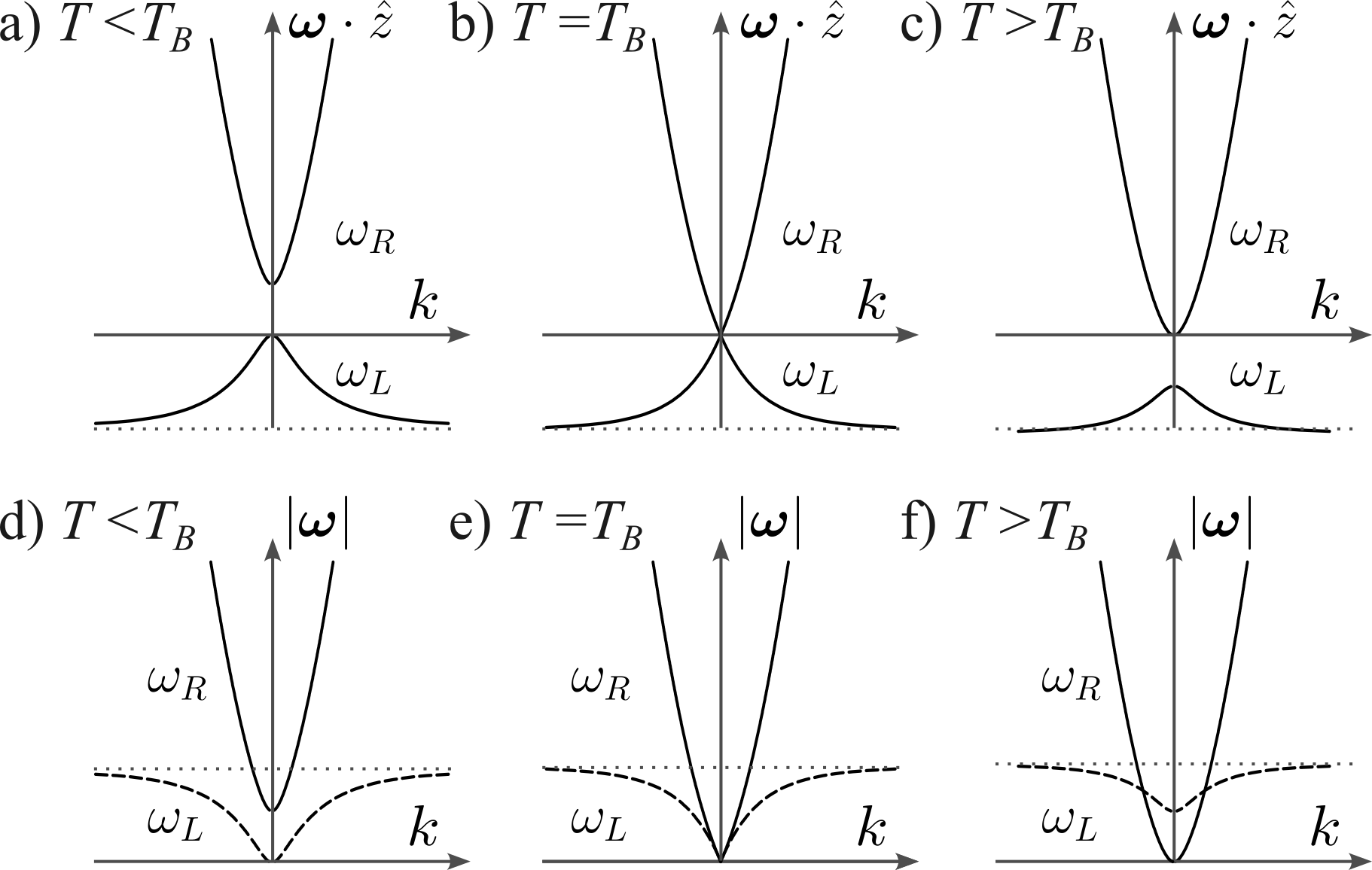}
\caption{Spin-wave spectrum of GdIG at temperatures (a,d) lower than, (b,e) equal to, and (c,f) higher than the compensation temperature $T_B$. Positive frequency ($\omega_R$) is associated with right-handed magnon. On the other hand, negative frequency ($\omega_L$), with a minimum frequency $\omega_\mathrm{min} \simeq -\lambda S^\mathrm{Fe}_z$, is associated with right-handed magnon branches. }
\label{FigSpectrum}
\end{figure}

The interaction leads to the following Landau–Lifshitz equations of motion for the spins
\begin{eqnarray}
\frac{d\textbf{S}^\mathrm{Fe}_{j}}{dt}&=&\textbf{S}^\mathrm{Fe}_{j}\times\left(\frac{I}{2}\left(\textbf{S}^\mathrm{Fe}_{j+1}+\textbf{S}^\mathrm{Fe}_{j-1}\right)-\lambda\textbf{S}^\mathrm{Gd}_{j}\right)\nonumber\\
\frac{d\textbf{S}^\mathrm{Gd}_{j}}{dt}&=&-\lambda\textbf{S}^\mathrm{Gd}_{j}\times\textbf{S}^\mathrm{Fe}_{j}.\label{EqLLG}
\end{eqnarray}
The spin-wave spectrum is evaluated by linearizing the Landau-Lifshitz equations using $F_{jx}\pm F_{jy}=F_\pm e^{i\left(jka-\omega t\right)}$
\begin{small}
\begin{eqnarray*}
&&i\frac{d}{dt}\left[
\begin{array}{c}
S^\mathrm{Fe}_+\\
S^\mathrm{Gd}_+
\end{array}
\right]=\left[
\begin{array}{cc}
IS^\mathrm{Fe}_z(1-\cos ka)-\lambda S^\mathrm{Gd}_z & \lambda S^\mathrm{Fe}_z\\
\lambda S^\mathrm{Gd}_z & -\lambda S^\mathrm{Fe}_z
\end{array}
\right]
\left[
\begin{array}{c}
S^\mathrm{Fe}_+\\
S^\mathrm{Gd}_+
\end{array}
\right],
\label{Eq.LinLLG}
\end{eqnarray*}
the eigen frequencies are
\begin{eqnarray*}
\omega= \frac{2D_\mathrm{YIG}}{a^2}\sin^2\frac{ka}{2}-\lambda\frac{S_z^\mathrm{Fe}+S_z^\mathrm{Gd}}{2}\\
\pm\sqrt{\left(\lambda\frac{S_z^\mathrm{Fe}+S_z^\mathrm{Gd}}{2}\right)^2-\frac{4\lambda D_\mathrm{YIG}S_z^\mathrm{Gd}}{a^2}\sin^2\frac{ka}{2}+\frac{4D^2_\mathrm{YIG}}{a^4}\sin^4\frac{ka}{2}}
\end{eqnarray*}
\end{small}
where $D_\mathrm{YIG}=Ia^2S^\mathrm{Fe}_z/2$ is the spin wave stiffness of YIG. Since $S^\mathrm{Gd}_z$ varies according to Eq.~\ref{Eq_Sgd}, the spectrum changes with temperatures, as illustrated in Fig.~\ref{FigSpectrum}. The positive and negative eigen frequencies are associated with right-handed (\textit{R}) and left-handed (\textit{L}) magnons, respectively. Their precession directions are illustrated in Fig.~\ref{Figantiferromagnon}.

\section{Results and Discussion}\label{sec:results}

$\mathcal{L}_0$ and $\mathcal{L}_1$ can be modified to account for the both magnon branches by substituting $S_z= S^\mathrm{Fe}_z+S^\mathrm{Gd}_z$ and
\begin{small}
\begin{eqnarray}
&&\int_{-\infty}^\infty \frac{\omega^n\mathrm{DOS}(\omega)d\omega}{e^{\beta\omega}-1}\nonumber\\
&&=
\int_{0}^\infty \frac{\omega_R^n\mathrm{DOS}(\omega_R)d\omega_R}{e^{\beta\omega_R}-1}
+\int_{-\infty}^0 \frac{\omega_L^n\mathrm{DOS}(|\boldsymbol\omega_L|)d|\boldsymbol\omega_L|}{e^{\beta|\boldsymbol\omega_L|}-1}\nonumber\\
\end{eqnarray}
\end{small}For negative energy spectra, one can see that the Bose-Einstein distribution become
\begin{equation}
\frac{1}{e^{\beta|\boldsymbol\omega_L|}-1}=\frac{1}{e^{-\beta\omega_L}-1}.
\end{equation}
Thus, the contributions of \textit{R} and \textit{L}-magnons are additive in $\mathcal{L}_0$
\begin{eqnarray*}
&\frac{\mathcal{L}_0}{2G\left|S_z^\mathrm{Fe}+S_z^\mathrm{Gd}\right|} = \int_{0}^\infty \frac{\mathrm{DOS}(\omega)d\omega}{e^{\beta\omega}-1} 
+\int_{-\infty}^0 \frac{\mathrm{DOS}(\omega)d\omega}{e^{-\beta\omega}-1} 
\end{eqnarray*}
but subtractive in $\mathcal{L}_1$
\begin{small}
\begin{eqnarray*}
&\frac{\mathcal{L}_1}{2G\left(S_z^\mathrm{Fe}+S_z^\mathrm{Gd}\right)}= \int_{0}^\infty \frac{\omega\mathrm{DOS}(\omega)d\omega}{e^{\beta\omega}-1} 
+\int_{-\infty}^0 \frac{\omega\mathrm{DOS}(\omega)d\omega}{e^{-\beta\omega}-1} 
\end{eqnarray*}
\end{small}
analogous to the transport of electrons and holes in semiconductors (see Table~\ref{TabComparison}). While band structure of semiconductor is weakly depends on temperature for as long as $k_BT\ll \epsilon_F$ is a lot smaller than the Fermi energy $\epsilon_F$, the magnon spectrum of ferrimagnetic Gd$_3$Fe$_5$O$_{12}$ is strongly depends on temperature, as illustrated in Fig.~\ref{FigSpectrum})

\begin{table}[h]

\label{TabComparison}
\caption{Comparison of thermal spin pumping from ferrimagnetic magnons to thermoelectric transport in semiconductor.}
\begin{tabular}{ccc}
\hline
\hline\\[-2ex]
Gd$_3$Fe$_5$O$_{12}$ & \ & Semiconductor\\
\\[-2ex]
\hline\\[-2ex]
Bosonic transport & & Fermionic transport
\\
$R$-magnon & & Electron
\\
($\omega_R$ branch) & & (conduction band)
\\
$n_R\propto \displaystyle\int \frac{\mathrm{DOS}(\omega_R)d\omega_R}{e^{\beta\omega_R}-1}$ 
& & $n_e\propto \displaystyle\int \frac{\mathrm{DOS}(\epsilon)d\epsilon}{e^{\beta(\epsilon-\epsilon_F)}+1}$
\\[2ex]
$L$-magnon & & Hole 
\\
($\omega_L$ branch) & & (valence band)
\\
$n_L\propto \displaystyle\int \frac{\mathrm{DOS}(\omega_L)d\omega_L}{e^{-\beta\omega_L}-1}$ 
& & $n_h\propto \displaystyle\int \frac{\mathrm{DOS}(\epsilon)d\epsilon}{e^{-\beta(\epsilon-\epsilon_F)}+1}$
\\
$\mathcal{L}_{0R}\mathcal{L}_{0L}>0$ & & $\mathcal{L}_{0e}\mathcal{L}_{0h}>0$
\\
$\mathcal{L}_{1R}\mathcal{L}_{1L}<0$ & & $\mathcal{L}_{1e}\mathcal{L}_{1h}<0$
\\
\hline
\hline
\end{tabular}
\end{table}

\subsection{High temperature spin pumping}
We first consider temperatures higher than the compensation temperature $T > T_B$. In this regime, the net magnetization $S^\mathrm{Fe}_z+S^\mathrm{Gd}_z$ is positive. Thus $\omega_R=\omega_1$ and $\omega_L=\omega_2$. For small $k$, the magnon branches are
\begin{eqnarray}
\omega_R&\simeq& 
\displaystyle\frac{S^\mathrm{Fe}_zD_\mathrm{YIG}k^2}{S^\mathrm{Fe}_z+S^\mathrm{Gd}_z}\nonumber\\
\omega_L&\simeq& 
-\lambda\left(S^\mathrm{Fe}_z+S^\mathrm{Gd}_z\right) +\displaystyle\frac{S^\mathrm{Gd}_zD_\mathrm{YIG}k^2}{S^\mathrm{Fe}_z+S^\mathrm{Gd}_z}.
\end{eqnarray}
\textit{R}-magnon dominate spin conductance at high temperatures because the number of \textit{L}-magnon remains significantly smaller.
\begin{eqnarray}
&\left(\frac{n_L}{n_R}\right)_{T>T_{B}}=\left(\frac{\mathcal{L}_{0L}}{\mathcal{L}_{0R}}\right)_{T>T_{B}}
=\frac{\int_{-\lambda S^\mathrm{Fe}_z}^{-\lambda\left(S^\mathrm{Fe}_z+S^\mathrm{Gd}_z\right)} \frac{\mathrm{DOS}\left(\omega_L\right)d\omega_L}{e^{-\beta\omega_L}-1}}{\int_0^{\infty} \frac{\mathrm{DOS}\left(\omega_R\right)d\omega_R}{e^{\beta\omega_R}-1}}
\nonumber\\
&\simeq \frac{4\left|\frac{S_z^\mathrm{Fe}}{S_z^\mathrm{Gd}}\right|^{\frac{3}{2}}}{\zeta\left(\frac{3}{2}\right)}
\int^{-\beta \lambda S^\mathrm{Gd}_z}_{0} \frac{(2\pi^2)^{-1}x^{\frac{1}{2}}dx}{e^{\beta\lambda\left(S^\mathrm{Fe}_z+S^\mathrm{Gd}_z\right)+x}-1} \ll 1.
\end{eqnarray} 
Here $\omega_\mathrm{min} \simeq -\lambda S^\mathrm{Fe}_z$ is the minimum frequency of the \textit{L}-magnon spectrum, as illustrated in Fig.~\ref{FigSpectrum}. Thus, the \textit{R}-magnon contribution dominates in high-temperature spin conductance.

\begin{figure}[t]
\includegraphics[width=\columnwidth]{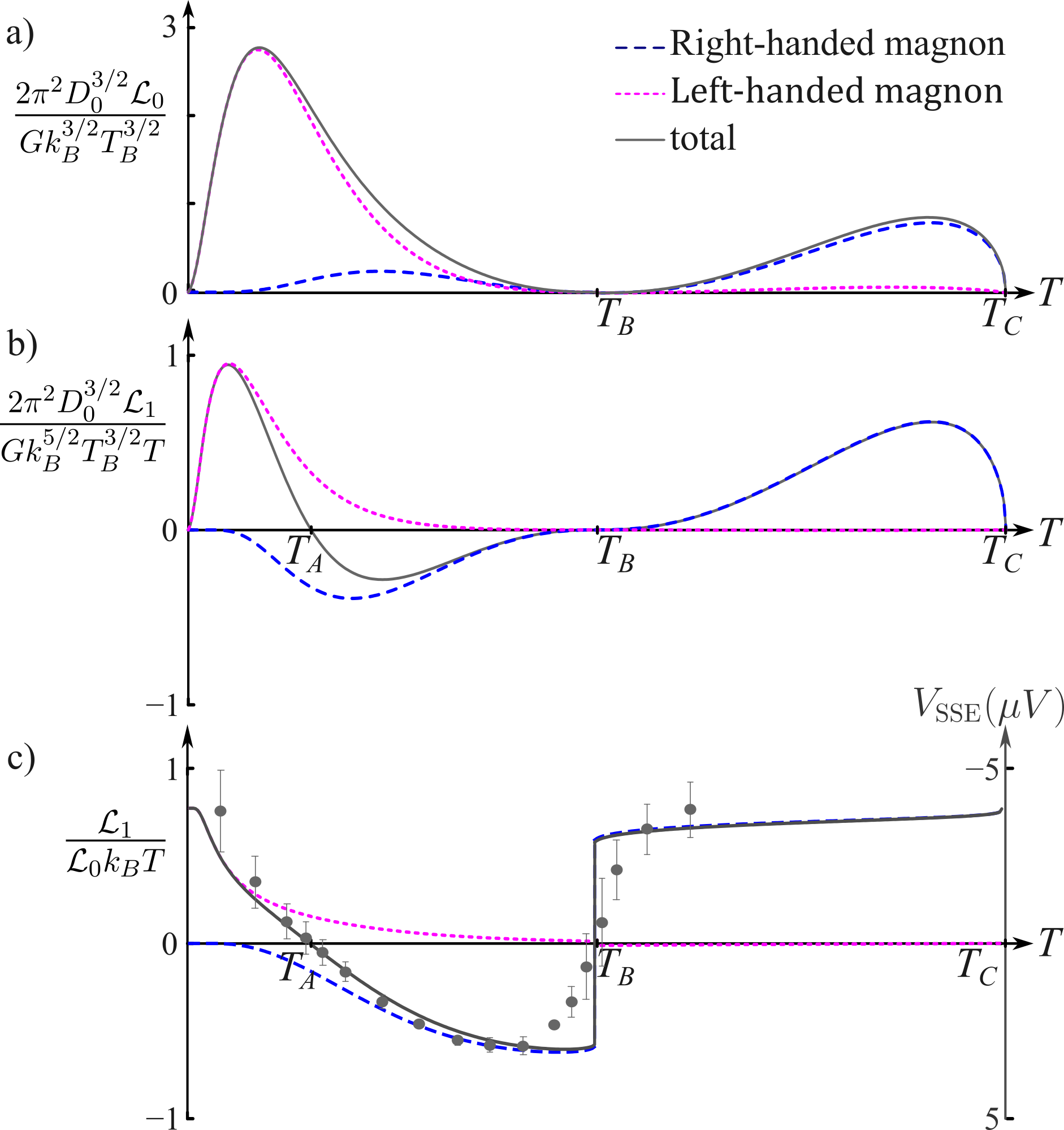}
\caption{(a) The interface spin injection conductance $\mathcal{L}_0$.\textit{R}-magnon dominates transport at high temperatures, while \textit{L}-magnon becomes dominant at low temperatures. (b) The thermal spin current $\mathcal{L}_1$ exhibits an additional compensation temperature $T_A$, where \textit{L}-magnon contribution exactly cancels \textit{R}-magnon contribution. (c) The ratio $\mathcal{L}_1/(\mathcal{L}_0T)$ determine the measured spin Seebeck voltage ($V_\mathrm{SSE}$). Our theoretical result is in agreement with experiment data from Ref.~\cite{Cramer2017}.}
\label{FigL01}
\end{figure}

In this temperature regime there is $|\boldsymbol\omega|$ spectrum crossing. However, the thermal spin pumping remains finite because the \textit{L}-magnon contribution is still much smaller
\begin{eqnarray}
&\left(\frac{\mathcal{L}_{1L}}{\mathcal{L}_{1R}}\right)_{T>T_{B}}
=\frac{\int_{-\lambda S^\mathrm{Fe}_z}^{-\lambda\left(S^\mathrm{Fe}_z+S^\mathrm{Gd}_z\right)} \frac{\omega_L\mathrm{DOS}\left(\omega_L\right)d\omega_L}{e^{-\beta\omega_L}-1}}{\int_0^{\infty} \frac{\omega_R\mathrm{DOS}\left(\omega_R\right)d\omega_R}{e^{\beta\omega_R}-1}}
\nonumber\\
&\simeq \frac{-8\left|\frac{S_z^\mathrm{Fe}}{S_z^\mathrm{Gd}}\right|^{\frac{3}{2}}}{3\zeta\left(\frac{5}{2}\right)}
\int^{-\beta \lambda S^\mathrm{Gd}_z}_{0} \frac{\left[\beta\lambda\left(S^\mathrm{Fe}_z+S^\mathrm{Gd}_z\right)+x\right]x^{\frac{1}{2}}dx}{2\pi^2\left[e^{\beta\lambda\left(S^\mathrm{Fe}_z+S^\mathrm{Gd}_z\right)+x}-1\right]}\ll 1.\nonumber\\
\end{eqnarray}
Thus, \textit{R}-magnon dominate $\mathcal{L}_1$ and the spin Seebeck coefficient $\mathcal{L}_1/(\mathcal{L}_0T)$ at high temperatures, as depicted in Fig.~\ref{FigL01}. 

\subsection{Low temperature spin pumping}
We now consider temperatures lower than the compensation temperature $T < T_B$, where $\omega_R=\omega_2$ and $\omega_L=\omega_1$. 
For small $k$, the magnon branches are
\begin{eqnarray}
\omega_R&\simeq& 
\lambda\left|S^\mathrm{Fe}_z+S^\mathrm{Gd}_z\right| -\frac{S^\mathrm{Gd}_zD_\mathrm{YIG}k^2}{\left|S^\mathrm{Fe}_z+S^\mathrm{Gd}_z\right|},\nonumber\\
\omega_L&\simeq& 
-\frac{S^\mathrm{Fe}_zD_\mathrm{YIG}k^2}{\left|S^\mathrm{Fe}_z+S^\mathrm{Gd}_z\right|}.
\end{eqnarray}
The magnon gap shifts to $\omega_R$. Thus, reducing number of \textit{R}-magnon 
\begin{eqnarray}
\left(\frac{n_L}{n_R}\right)_{T<T_{B}}&=&\left(\frac{\mathcal{L}_{0L}}{\mathcal{L}_{0R}}\right)_{T<T_{B}}=\frac{\int_{-\lambda S^\mathrm{Fe}_z}^0 \frac{\mathrm{DOS}(\omega_L)d\omega_L}{e^{-\beta\omega_L}-1}}{\int_{\lambda\left|S^\mathrm{Fe}_z+S^\mathrm{Gd}_z\right|}^{\infty} \frac{\mathrm{DOS}(\omega_R)d\omega_R}{e^{\beta\omega_R}-1}}
\nonumber\\
&=& \frac{\frac{2}{\sqrt{\pi}}\left|\frac{S_z^\mathrm{Gd}}{S_z^\mathrm{Fe}}\right|^{\frac{3}{2}}}{\mathrm{Li}_{\frac{3}{2}}\left(e^{-\beta\lambda \left|S^\mathrm{Fe}_z+S^\mathrm{Gd}_z\right)}\right|}
\int^{\beta\lambda S^\mathrm{Fe}_z}_{0} \frac{x^{\frac{1}{2}}dx}{e^{x}-1}
\end{eqnarray}
As the temperature decreases, the \textit{L}-magnon population increases, making it the dominant contribution at low temperatures. Figure~\ref{FigL01}a shows the variation of $\mathcal{L}_0$ with temperature. At low temperatures, \textit{L}-magnon dominate the spin conductance, with the asymptotic low-temperature limit
\begin{eqnarray}
\lim_{T\to 0}\mathcal{L}_0=\frac{G\zeta\left(\frac{3}{2}\right)\left(k_BT\right)^{{3}/{2}}\left|S^\mathrm{Fe}_z+S^\mathrm{Gd}_z\right|^{{5}/{2}}}{2\left(\pi D_0S^\mathrm{Fe}_z\right)^{{3}/{2}}}.
\end{eqnarray}
Similarly, the \textit{R}-magnon contribution to $\mathcal{L}_1$ is exponentially suppressed 
\begin{eqnarray}
&&\left(\frac{\mathcal{L}_{1L}}{\mathcal{L}_{1R}}\right)_{T<T_{B}}
=\frac{\int_{-\lambda S^\mathrm{Fe}_z}^0 \frac{\omega_L\mathrm{DOS}(\omega_L)d\omega_L}{e^{-\beta\omega_L}-1}}{\int_{-\lambda\left|S^\mathrm{Fe}_z+S^\mathrm{Gd}_z\right|}^{\infty} \frac{\omega_R\mathrm{DOS}(\omega_R)d\omega_R}{e^{\beta\omega_R}-1}}
\nonumber\\
&&\simeq \frac{-\frac{4}{\sqrt{\pi}}\left|\frac{S_z^\mathrm{Gd}}{S_z^\mathrm{Fe}}\right|^{\frac{3}{2}}}{3\mathrm{Li}_{\frac{5}{2}}\left(e^{-x_0}\right)-2x_0\mathrm{Li}_{\frac{3}{2}}\left(e^{-x_0}\right)}
\int^{\beta\lambda S^\mathrm{Fe}_z}_{0} \frac{x^{\frac{3}{2}}dx}{e^{x}-1}.
\end{eqnarray}
Here $x_0=\beta\lambda\left|S^\mathrm{Fe}_z+S^\mathrm{Gd}_z\right|$. The asymptote value for low temperature is
\begin{eqnarray}
\lim_{T\to 0}\mathcal{L}_1
=\frac{3G\zeta\left(\frac{5}{2}\right)\left(k_BT\right)^{\frac{5}{2}}\left|S^\mathrm{Fe}_z+S^\mathrm{Gd}_z\right|^{\frac{5}{2}}}{4\pi^{\frac{3}{2}} D_0^{\frac{3}{2}}\left(S^\mathrm{Fe}_z\right)^{\frac{3}{2}}}.
\end{eqnarray}
 
Although there is no $|\boldsymbol\omega|$ crossing when $T<T_B$, a compensation temperature $T_A$ occurs when the total $\mathcal{L}_{1R}$ and $\mathcal{L}_{1L}$ cancel each others, as illustrated in Fig.\ref{FigL01}b.  At $T_A$, the thermal spin current carried by right-handed and left-handed magnons cancel each other. Reciprocally, the net heat current carried by both magnon modes in under finite spin accumulation is also zero due to Onsager relation in matriks Eq.~\ref{EqMatriksSpinHeat}. The value of $\mathcal{L}_1$ becomes negative for $T_A<T<T_B$.
Numerical evaluation of the integrals yields $T_A/T_B=0.3$, which aligns well with experimental values reported in Ref.~\cite{Geprags2016} ($T_A=80$ K and $T_B=256.5$ K) and Ref.~\cite{Cramer2017} ($T_A=72$ K and $T_B\simeq 250$ K).

Our findings contrast with the interpretation presented in Ref.~\cite{Li2022}, where the compensation temperature is attributed to a spectral crossing of $|\boldsymbol\omega_R|$ and $|\boldsymbol\omega_L|$ obtained from a full multi-sublattice exchange model. In that framework, the vanishing of the spin current is interpreted as a direct consequence of spectral symmetry. In our model, however, we show that such a spectral crossing is neither a necessary nor sufficient condition for compensation. Instead, we identify the compensation temperature $T_A$ as arising from a precise cancellation between the thermally weighted contributions of right- and left-handed magnons, governed by their distinct frequency branches and temperature-dependent populations. Our results therefore offer an alternative interpretation of the origin of compensation and highlight the role of magnon statistics rather than spectral degeneracy alone. This distinction provides theoretical clarity and suggests that compensation may occur under broader conditions than previously assumed.

\section{Conclusion}\label{sec:conclusion}
In this work, we investigated the mechanisms of thermal spin pumping in gadolinium iron garnet (GdIG) by analyzing the distinct contributions of left-handed and right-handed antiferromagnetic magnons. Using a two-magnetic-lattice model with simplified Fe–Fe and Fe–Gd exchange interactions, we derived the temperature-dependent spin-wave spectrum and evaluated the corresponding spin transport coefficients $\mathcal{L}_0$ and $\mathcal{L}_1$.

We treated right-handed and left-handed magnons as positive and negative frequency branches, respectively. This approach is conceptually analogous to electron and hole transport in semiconductors, where carriers in opposite bands contribute with opposite signs to current. At high temperatures, right-handed magnons dominate the thermal spin current due to their higher population. As the temperature decreases, their contribution becomes suppressed due to the opening of a magnon gap, while left-handed magnons emerge as the primary carriers.

This competition leads to a compensation temperature, at which the thermally driven spin current vanishes. Notably, we found that this compensation point does not coincide with the crossing of $|\boldsymbol\omega_R|$ and $|\boldsymbol\omega_L|$ spectra. Furthermore, we observed that spin pumping remains finite at temperatures where such crossings occur. This behavior highlights the fact that compensation results from the integrated statistical contributions of each mode, rather than from spectral symmetry alone.

Our findings provide an alternative interpretation to that of Ref.~\cite{Li2022}, which attributes compensation to a spectral crossing in a detailed multi-sublattice exchange model. In contrast, our simplified two-sublattice framework captures key features of spin pumping behavior in a more analytically transparent form. This model offers both qualitative and quantitative insights into mode-resolved spin transport in ferrimagnets. 

The theoretical framework developed in this work contributes a physically intuitive and computationally efficient tool for analyzing thermal spin pumping. It is well suited for guiding experimental interpretation and for informing the design of spintronic devices based on chiral magnon dynamics and thermally driven spin currents.

\textit{Acknowledgments.} 
We acknowledge the support of Universitas Indonesia via PUTI Research Grant No. NKB-366/UN2.RST/HKP.05.00/2024

\appendix

\section{Gd-Gd exchange contribution to compensation temperatures}
\label{SecAppendix}
Here we estimate the contribution of Gd-Gd exchange, characterized by $I_\mathrm{Gd}\ll \lambda\ll I$, to the compensation temperatures using the following Hamiltonian that includes magnetic field 
\begin{eqnarray*}
\mathcal{H}=&- \mu_0 \textbf{H}\cdot 2\mu_B\left(\textbf{S}^\mathrm{Fe}+\textbf{S}^\mathrm{Gd}\right) -\frac{I}{2}\sum_j \textbf{S}^\mathrm{Fe}_{j}\cdot\textbf{S}^\mathrm{Fe}_{j+1} \nonumber\\
&+\lambda\sum_{j} \textbf{S}^\mathrm{Gd}_{j}\cdot\textbf{S}^\mathrm{Fe}_n-\frac{I_\mathrm{Gd}}{2}\sum_j \textbf{S}^\mathrm{Gd}_{j}\cdot\textbf{S}^\mathrm{Gd}_{j+1} , 
\end{eqnarray*}
where $\mu_B$ is the Bohr magneton. The statistical average is as follows
\begin{eqnarray*}
S^\mathrm{Gd}_z&=&3S^\mathrm{Gd}B_{S^\mathrm{Gd}}\left(S^\mathrm{Gd}\frac{2\mu_B \mu_0 H+I_\mathrm{Gd}S^\mathrm{Gd}_z-\lambda S^\mathrm{Fe}_z}{k_BT}\right)
\nonumber\\
S^\mathrm{Fe}_z&=&S^\mathrm{Fe}B_{S^\mathrm{Fe}}\left(S^\mathrm{Fe}\frac{2\mu_B \mu_0 H+I S^\mathrm{Fe}_z-\lambda S^\mathrm{Gd}_z}{k_BT}\right).
\end{eqnarray*}

\subsection{Contribution on $T_C$}
For high temperatures, $S^\mathrm{Gd}_z$ and $S^\mathrm{Fe}_z$ can be linearised as follows
\begin{eqnarray*}
\left[
\begin{array}{cc}
1-\frac{I^\mathrm{Gd}S^\mathrm{Gd}\left(S^\mathrm{Gd}+1\right)}{k_BT} & \frac{\lambda}{k_BT}\\
\frac{\lambda}{k_BT} & 1-\frac{IS^\mathrm{Fe}\left(S^\mathrm{Fe}+1\right)}{3k_BT}
\end{array}
\right]
\left[
\begin{array}{c}
S^\mathrm{Gd}_z\\
S^\mathrm{Fe}_z
\end{array}
\right]\\
\approx
\left[
\begin{array}{c}
\frac{2\mu_B\mu_0 HS^\mathrm{Gd}\left(S^\mathrm{Gd}+1\right)}{k_BT}\\
\frac{2\mu_B\mu_0 HS^\mathrm{Fe}\left(S^\mathrm{Fe}+1\right)}{3k_BT}
\end{array}
\right]
\end{eqnarray*}
The Curie temperature occurred when the determinant of the matrix  is zero.
\begin{small}
\begin{eqnarray*}
T_C
&\approx&\frac{IS^\mathrm{Fe}\left(S^\mathrm{Fe}+1\right)}{3k_B}+\frac{3\lambda^2}{k_BIS^\mathrm{Fe}\left(S^\mathrm{Fe}+1\right)}\left(1+\frac{3I_\mathrm{Gd}}{I}\right)
\end{eqnarray*}
\end{small}

\subsection{Contribution on $T_B$}
$S^\mathrm{Gd}_z$ on zero magnetic field can be linearized as follows
\begin{eqnarray*}
S^\mathrm{Gd}_z(T)\approx S^\mathrm{Gd}\left(S^\mathrm{Gd}+1\right)\frac{I_\mathrm{Gd}S^\mathrm{Gd}_z-\lambda S^\mathrm{Fe}_z}{k_BT}
\\
\approx -\lambda S^\mathrm{Fe}_z\frac{S^\mathrm{Gd}\left(S^\mathrm{Gd}+1\right)}{k_BT}\left(1+\frac{I_\mathrm{Gd}S^\mathrm{Gd}\left(S^\mathrm{Gd}+1\right)}{k_BT}\right).
\end{eqnarray*} 
The modified compensation temperature $T_B$ as follows
\begin{eqnarray*}
0&=&S^\mathrm{Gd}_z(T_B)+S^\mathrm{Fe}_z\\
0&=&1-\lambda \frac{S^\mathrm{Gd}\left(S^\mathrm{Gd}+1\right)}{k_BT_B}\left(1+\frac{I_\mathrm{Gd}S^\mathrm{Gd}\left(S^\mathrm{Gd}+1\right)}{k_BT}\right)
\end{eqnarray*}
such that
\begin{eqnarray*}
T_B
\approx \frac{\lambda S^\mathrm{Gd}\left(S^\mathrm{Gd}+1\right)}{k_B}\left(1+\frac{I_\mathrm{Gd}}{\lambda}\right)
\end{eqnarray*}

\subsection{Contribution on $T_A$}
The Landau–Lifshitz equations of motion for the spins become
\begin{eqnarray*}
\frac{d\textbf{S}^\mathrm{Fe}_{j}}{dt}&=&\textbf{S}^\mathrm{Fe}_j\times\left(\frac{I}{2}\left(\textbf{S}^\mathrm{Fe}_{j+1}+\textbf{S}^\mathrm{Fe}_{j-1}\right)-\lambda\textbf{S}^\mathrm{Gd}_{j}\right)\nonumber\\
\frac{d\textbf{S}^\mathrm{Gd}_{j}}{dt}&=&\textbf{S}^\mathrm{Gd}_{j}\times\left(-\lambda\textbf{S}^\mathrm{Fe}_{j}+\frac{I_\mathrm{Gd}}{2}\left(\textbf{S}^\mathrm{Gd}_{j+1}+\textbf{S}^\mathrm{Gd}_{j-1}\right)\right).
\end{eqnarray*}
The spin-wave spectrum for small $k$ is
\begin{eqnarray*}
i\frac{d}{dt}\left[
\begin{array}{c}
S^\mathrm{Fe}_+\\
S^\mathrm{Gd}_+
\end{array}
\right]=W
\left[
\begin{array}{c}
S^\mathrm{Fe}_+\\
S^\mathrm{Gd}_+
\end{array}
\right],\\
W=\left[
\begin{array}{cc}
IS^\mathrm{Fe}_z\frac{(ka)^2}{2}-\lambda S^\mathrm{Gd}_z & \lambda S^\mathrm{Fe}_z\\
\lambda S^\mathrm{Gd}_z & I_\mathrm{Gd}S^\mathrm{Gd}_z\frac{(ka)^2}{2}-\lambda S^\mathrm{Fe}_z
\end{array}
\right]
\end{eqnarray*}
the dispersion relation are
\begin{eqnarray*}
\frac{k^2a^2}{2}=\frac{\omega\left(\omega+\lambda\left(S_z^\mathrm{Fe}+S_z^\mathrm{Gd}\right)\right)}{\left(IS_z^\mathrm{Fe}+I_\mathrm{Gd}S_z^\mathrm{Gd}\right)+\lambda\left(I\left(S_z^\mathrm{Fe}\right)^2+I_\mathrm{Gd}\left(S_z^\mathrm{Gd}\right)^2\right)}
\end{eqnarray*}
The minimum frequency is now $$\omega_\mathrm{min} \simeq -\lambda S^\mathrm{Fe}_z\left(1+\frac{I_\mathrm{Gd}S_z^\mathrm{Gd}}{IS_z^\mathrm{Fe}}\left(\frac{S_z^\mathrm{Gd}}{S_z^\mathrm{Fe}}-1\right)\right).$$
$I_\mathrm{Gd}$ modifies the spin wave dispersion, which in turn determine $T_A$. For $T>T_B$,
\begin{eqnarray*}
\omega_R&\approx& 
\displaystyle\frac{S^\mathrm{Fe}_zD_\mathrm{YIG}k^2}{S^\mathrm{Fe}_z+S^\mathrm{Gd}_z}\left(1+\frac{I_\mathrm{Gd}}{I}\left(\frac{S^\mathrm{Gd}_z}{S^\mathrm{Fe}_z}\right)^2\right)\\
\omega_L&\simeq& 
-\lambda\left(S^\mathrm{Fe}_z+S^\mathrm{Gd}_z\right) +\displaystyle\frac{S^\mathrm{Gd}_zD_\mathrm{YIG}k^2}{S^\mathrm{Fe}_z+S^\mathrm{Gd}_z}\left(1+\frac{I_\mathrm{Gd}}{I}\right).
\end{eqnarray*}
Meanwhile, for $T<T_B$,
\begin{eqnarray*}
\omega_R&\approx& 
\lambda\left|S^\mathrm{Fe}_z+S^\mathrm{Gd}_z\right| -\frac{S^\mathrm{Gd}_zD_\mathrm{YIG}k^2}{\left|S^\mathrm{Fe}_z+S^\mathrm{Gd}_z\right|}\left(1+\frac{I_\mathrm{Gd}}{I}\right),\nonumber\\
\omega_L&\approx& 
-\frac{S^\mathrm{Fe}_zD_\mathrm{YIG}k^2}{\left|S^\mathrm{Fe}_z+S^\mathrm{Gd}_z\right|}\left(1+\frac{I_\mathrm{Gd}}{I}\left(\frac{S^\mathrm{Gd}_z}{S^\mathrm{Fe}_z}\right)^2\right).
\end{eqnarray*}

\section*{References}

\end{document}